\pdfoutput=1
\documentclass[final,3p,times,twocolumn,sort&compress]{elsarticle}
\usepackage{graphicx}
\usepackage{amssymb}
\usepackage{amsmath}
\usepackage{xcolor}
\definecolor{blue}{cmyk}{0.68,0.66,0,0.43}
\usepackage{float}
\usepackage{microtype}
\usepackage[pdfauthor={Thomas Mueller, Joost de Groot, Joerg Strempfer, Manuel Angst},
            pdftitle={Stoichiometric YFe2O4-delta single crystals grown by the optical floating zone method.},
            pdfsubject={},
            pdfkeywords={YFe2O4, LuFe2O4, Rare earth compounds, Floating zone technique, X-ray diffraction, Charge Ordering},
            pdfproducer={Latex with hyperref},
            pdfcreator={pdflatex}]{hyperref}
\hypersetup{colorlinks=true,allcolors=green}
\journal{Journal of Crystal Growth}
\usepackage{nicefrac}
\usepackage{caption}
\usepackage[labelformat=simple]{subcaption}

\usepackage[absolute]{textpos}

\begin{document}

    \begin{textblock}{14}(1.5,14.5)    
         \noindent
         \footnotesize
        \copyright 2015. This manuscript version is made available under the CC-BY-NC-ND 4.0 license \url{http://creativecommons.org/licenses/by-nc-nd/4.0/}
    \end{textblock}

\begin{frontmatter}
\title{Stoichiometric YFe$_2$O$_{4-\delta}$ single crystals  \\ grown by the optical floating zone method.}
\author[1]{Thomas Mueller\corref{cortm}}
\ead{thomas.mueller6@rwth-aachen.de +492461613324}
\author[1]{Joost de Groot}
\author[2]{J\"org Strempfer}
\author[1,3]{Manuel Angst\corref{cortm}}
\ead{m.angst@fz-juelich.de}
\address[1]{J\"ulich Centre for Neutron Science JCNS and Peter Gr\"unberg Institut PGI, JARA-FIT, Forschungszentrum J\"ulich GmbH, 52425 J\"ulich, Germany.}
\address[3]{Experimental Physics IVC, RWTH Aachen University, 52056 Aachen, Germany.}
\address[2]{Deutsches Elektronen-Synchrotron (DESY), 22603 Hamburg, Germany.}
\cortext[cortm]{Corresponding author}

\begin{abstract}
We report the growth of YFe$_2$O$_{4-\delta}$ single crystals by the optical floating zone method, showing for the first time the same magnetization as highly stoichiometric $(\delta = 0.00)$ powder samples and sharp superstructure reflections in single crystal x-ray diffraction. The latter can be attributed to three dimensional long-range charge ordering. Resonant x-ray diffraction at the Fe K-edge with full linear polarization analysis was used for the investigation of the possibility of orbital order.
\end{abstract}

\end{frontmatter}

\section{Introduction}
\label{sec:intro}
LuFe$_2$O$_{4-\delta}$ is still in debate as a multiferroic compound with ferroelectricity due to charge ordering at room temperature~\cite{Ikeda2005}. The ferroelectricity was strongly questioned~\cite{joost_co,Niermann2012,Ruff2012,Lafuerza2013}, but was again picked up recently~\cite{Kambe2013}. Nevertheless, the mechanism could also be applicable to other rare earth ferrites.
There is much less known about other members of the $R$Fe$_2$O$_4$ family, where $R$ denotes rare earths, for example Y, Ho, Er or Yb. For recent reviews see~\cite{Angst2013,Ikeda2015}. These materials crystallize in a rhombohedral structure with space group R$\bar{3}$m~\cite{Kimizuka1974,Kimizuka1975,ISOBE1990}. Triangular layers of the rare earth ions alternate with two triangular layers of Fe, stacked along the hexagonal $\vec{c}$ axis (Fig.~\ref{fig:cells}). The oxygen ions form trigonal bipyramidal environments for the Fe ions and octahedral ones for the Yttrium. The average Fe valence of 2.5 leads to an arrangement of bi- and trivalent Fe-ions on a triangular lattice. The frustration intrinsic to the triangular lattices gives rise to complex ordering phenomena \cite{Angst2013}.

In the $R$Fe$_2$O$_4$ family YFe$_2$O$_{4-\delta}$ is of particular interest, because Y$^{3+}$ has the largest ionic radius and strong differences compared to LuFe$_2$O$_4$ are observed in electron diffraction \cite{Matsuo2010,Mori2008}. Oxygen-stoichiometry has a huge influence on the occurrence of  three dimensional long range charge and spin order in YFe$_2$O$_{4-\delta}$ \cite{Funahashi1984,Mori2008,Horibe2009}. 
All previous work on stoichiometric YFe$_2$O$_{4-\delta}$ was based on powder samples (e.g.\ \cite{Katano1995,Funahashi1984}) and electron diffraction on tiny crystallites obtained from powder (e.g.\ \cite{Horibe2010,Horibe2009,Mori2008}). Stoichiometric single crystals, large enough for single crystal x-ray diffraction,  were not grown due to decomposition problems during growth. This decomposition increases at higher oxygen partial pressure \cite{Shindo1976}.
We report the growth of YFe$_2$O$_{4-\delta}$ single crystals by the optical floating zone method. Our crystals for the first time show the same macroscopic magnetization as highly stoichiometric powder samples $(\delta=0.00)$~\cite{Inazumi1981}. They exhibit sharp superstructure reflections in x-ray diffraction and are large enough (up to 52\,mg) for neutron diffraction.

\section{Crystal Growth} \label{sec:crystalgrowth}
\begin{figure}[tb]
\includegraphics{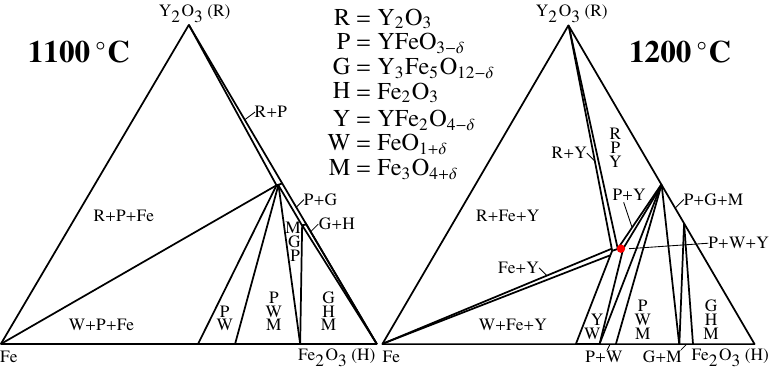}
\caption{Phase diagram of the Y$_2$O$_3$-Fe-Fe$_2$O$_3$ System at 1100\,$^\circ\textrm{C}$ \cite{Kitayama2004}  and 1200\,$^\circ\textrm{C}$ \cite{Inazumi1981}. The region where stoichiometric YFe$_2$O$_{4-\delta}$ exists, is marked in \textcolor{red}{red} in the 1200\,$^\circ\textrm{C}$ diagram. (color online)}
\label{fig:phase}
\end{figure}

\begin{figure}[tb]
\setlength\abovecaptionskip{0pt}
\setlength\belowcaptionskip{0pt}
\includegraphics[width=215px]{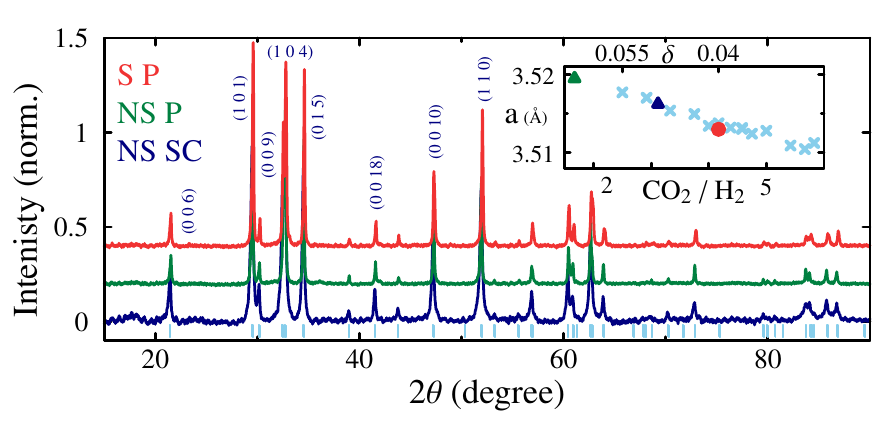}
\caption{Powder diffractograms of YFe$_2$O$_{4-\delta}$ for the non-stoichiometric single crystal (\textcolor[HTML]{000080}{NS SC}) with the magnetization in the inset of Fig.~\ref{fig:mag}, non-stoichiometric powder $(\delta > 0.05)$ (\textcolor[HTML]{008040}{NS P}) and almost stoichiometric powder $(0.03< \delta <0.04)$ (\textcolor[HTML]{F03232}{S P}), with the two step transition as in Fig.~\ref{fig:mag} but a difference at low temperatures between FC and ZFC. In light blue the reflection positions for the R$\bar{3}$m structure reported in \cite{Matsumoto1992a} are given.
  \\ \textit{Inset}: Dependency of the R$\bar{3}$m a/b-lattice parameter on the synthesis gas mixture. The given $\delta$ is determined from comparison of the magnetization data with those from \cite{Inazumi1981}, the correlation to the gas ratio is nonlinear \cite{Jacob2012}. For the single crystal the gas ratio was interpolated from the lattice parameter and the powder data, since it was synthesized in a CO$_2$~/~CO mixture. 
  (color online)}
\label{fig:powderdiff}
\end{figure}

Following Shindo~\textit{et\,al.\ }\cite{Shindo1976}, powders of Fe$_2$O$_3$ (99.945\%) and Y$_2$O$_3$ (99.99\%) were mixed in the stoichiometric ratio. Afterwards, they were calcined at 1250\,$^\circ$C for 24\,h in a CO$_2$/Ar-H$_2$(4\%) atmosphere to control the oxygen partial pressure. According to \cite{Kimizuka1975} YFe$_2$O$_{4-\delta}$ is present as a stable phase at 1200\,$^\circ$C in the  Y-Fe-O-system (Fig.~\ref{fig:phase}) with a stoichiometry ranging from $\delta= 0.095$ to $\delta=0.00$.  Below 1100\,$^\circ$C YFe$_2$O$_{4-\delta}$ is only metastable~\cite{Kitayama2004}. To preserve the metastable state, the crucible containing the material was moved manually from the hot zone to the room temperature region of the tube furnace and was cooled inside the same atmosphere for 1\,h.  The resulting polycrystalline YFe$_2$O$_{4-\delta}$ was ground and hydrostatically pressed at 30\,MPa into rods of 7\,mm diameter and typically 70\,mm length. Afterwards the rods were sintered in the same low oxygen atmosphere. These polycrystalline rods were used for both the feed and the seed. The oxygen partial pressure was fine tuned by analyzing samples grown under different CO$_2$/H$_2$-ratios by powder x-ray diffraction and magnetization measurements. We chose a CO$_2$/H$_2$-ratio of 3 for the rod preparation. This gave non-stoichiometric YFe$_2$O$_{4-\delta}$, but with the use of higher oxygen partial pressure the occurrence of foreign phases was more likely. The final stoichiometry of the crystals is determined by the atmosphere during crystal growth. Figure~\ref{fig:powderdiff} shows the powder diffractograms of two polycrystalline samples and a ground non-stoichiometric $(\delta > 0.05)$ single crystal. The inset shows the dependency of the $a$ lattice parameter on the synthesis gas ratios. The corresponding $c$ lattice parameters range from 24.746 to 24.765 but no general trend is visible. The error on the $c$ lattice parameter is mostly determined by the reproducibility of powders with a specific stoichiometry. For powders synthesized under the same conditions the standard deviation of the $c$ lattice parameter is around 0.007. The overall unit cell volume decreases slightly with increasing oxygen content (by 0.4\% upon changing the CO$_2$/ H$_2$ ratio from 1.7 to 5.6), it is mainly caused by a decrease of the a\,/\,b lattice parameter (see inset Fig.\ref{fig:powderdiff}). This is expected \cite{Conder2001,nemundr} and also observed in LuFe$_2$O$_{4-\delta}$ \cite{Bourgeois2012,Sekine1976}. However, it is in contradiction to \cite{Inazumi1981}, where a small ($<0.1$\%) volume increase is observed. 

An optical four mirror furnace (Model FZ-T-10000-H-VI-VP0 Crystal Systems Corporation, Japan) was used for the floating zone growth. The atmosphere was controlled with a mixture of CO$_2$ and CO. Using CO$_2$ and H$_2$ would create water, which disturbs the growth \cite{IIDA1990}. Growth rates below 2\,\nicefrac{mm}{h} have been found essential to grow YFe$_2$O$_{4-\delta}$ \cite{Shindo1976}. We used a growth rate of 1\,\nicefrac{mm}{h} and seed and feed rods counter-rotating with 10 and 20\,rpm, respectively. The grown crystal slowly cools to room temperature, while moving out of the lamp focus. This leads to a partial decomposition of YFe$_2$O$_{4-\delta}$ to polycrystalline perovskite YFeO$_{3-\delta}$ and FeO$_{1+\delta}$, as checked by powder x-ray diffraction and already observed in \cite{Shindo1976}. The use of a crystalline seed was attempted, but was not beneficial, because the metastable YFe$_2$O$_{4-\delta}$ starts to decompose before it melts.

\begin{figure}[tb]
\setlength\abovecaptionskip{0pt}
\setlength\belowcaptionskip{0pt}
\includegraphics[width=220px]{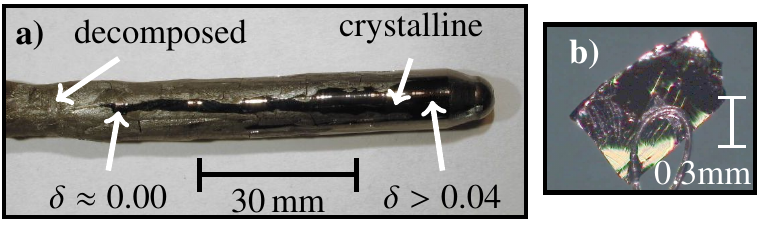}
{\phantomsubcaption\label{fig:boulea}}
{\phantomsubcaption\label{fig:crystalsn}}
\caption{a) Grown boule, consisting of crystalline YFe$_2$O$_{4-\delta}$ and material decomposed to YFeO$_3$ and FeO during cooling. The last grown part (right side) was quenched, through the separation of the seed and feed rod at the end of the growth. Strong variations are observed in the oxygen off-stoichiometry $\delta$.  The given oxygen off-stoichiometry $\delta$ is determined from comparison of the magnetization with~\cite{Inazumi1981}. \\ b) Tiny separated single crystal on a nylon string used for single crystal x-ray diffraction. (color online) }
\end{figure}

The grown rod (Fig.~\ref{fig:boulea}) was crushed to separate the YFe$_2$O$_{4-\delta}$ from the decomposed parts, which can be easily distinguished by optical microscopy.  Despite the layered structure, YFe$_2$O$_{4-\delta}$ does not prefer to cleave in any specific orientation, unlike LuFe$_2$O$_{4-\delta}$~\cite{joostphd}, which easily cleaves at facets perpendicular  to $\vec{c}_\textrm{\tiny{hex}}$.

The oxygen-stoichiometry of YFe$_2$O$_{4-\delta}$ can be tuned by the oxygen partial pressure during growth \cite{Jacob2012,Inazumi1981}. Stoichiometric single crystals $(\delta <0.03)$ could only be grown with a CO$_2$~/~CO-ratio of 2.9$\pm$0.1. This is significantly higher than the ratio of 0.4 used by Shindo \textit{et\,al.} to grow non-stoichiometric YFe$_2$O$_{4-\delta}$~\cite{Shindo1976}, but it is much closer to the ratio of 2.7 used for stoichiometric LuFe$_2$O$_4$ \cite{Christianson2008}. Surprisingly, also the position in the grown rod has a huge influence on stoichiometry. The last grown part tends to be non-stoichiometric. 
The decomposition process is prevented in this part due to the faster cooling. The only stoichiometric samples we could obtain are from the first grown not decomposed part (see Fig.~\ref{fig:boulea}).
This could have several reasons. One possibility would be that the composition of the melt changes during the growth, which would be compatible with the theory of incongruently melting YFe$_2$O$_{4-\delta}$~\cite{Shindo1976}. This possibility seems unlikely, because the stoichiometry is independent of the overall growth length. If the growth is stopped after the first crystalline part has grown, this part is non-stoichiometric.

That leaves the possibility of a change from non-stoichiometric to stoichiometric YFe$_2$O$_{4-\delta}$ during cooling. The oxygen partial pressure changes with the temperature \cite{trevor} and also the oxygen stoichiometry of the solid phase can shift the equilibrium of the gas mixture significantly \cite{Darken1945},  which could cause annealing of the grown part. However with decreasing temperature the equilibrium between CO$_2$ and CO is shifted to the CO$_2$ side, leading to less free oxygen.  Another possible influence is the decomposition process to YFeO$_3$ and FeO. While the process with stoichiometric compounds is neutral in regard to free oxygen, the formation of FeO$_{1-\delta}$ or YFeO$_{3-\delta}$ with positive $\delta$ could supply additional oxygen to the YFe$_2$O$_{4-\delta}$ crystals. FeO$_{1+\delta}$ prefers the iron deficient form, even in a reducing atmosphere \cite{Darken1945,Jette1933}. While the formation of YFeO$_{3-\delta}$ is common \cite{Jacob2012,Kitayama2004}, the lower phase stability limit $\delta=0.03$ \cite{Kitayama2004} for YFeO$_{3-\delta}$ is in the same region as the minimal oxygen excess in FeO$_{1+\delta}$ \cite{Smyth1961167}. This makes an oxygen provision to YFe$_2$O$_{4-\delta}$ from the decomposition process possible.

\section{Macroscopic characterization}
\begin{figure}[tb]
\setlength\abovecaptionskip{0pt}
\setlength\belowcaptionskip{0pt}
\includegraphics{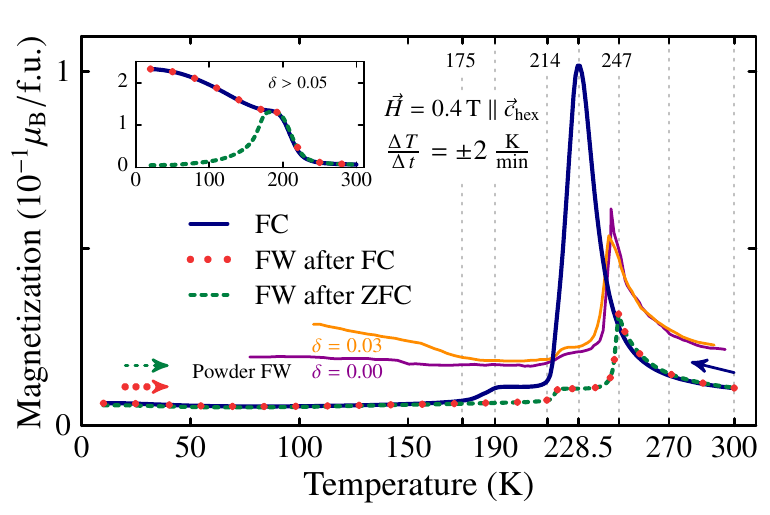}
\caption{Magnetization of a 52\,mg \textit{stoichiometric}  YFe$_2$O$_{4-\delta}$ single crystal $(0.00 \approx \delta \ll  0.03)$ grown in a CO$_2$~/~CO-ratio of 2.9$\pm$0.1, measured during field cooling (\textcolor[HTML]{000080}{FC}), field warming (\textcolor[HTML]{F03232}{FW}) after cooling in a field and after cooling without a field (\textcolor[HTML]{008040}{ZFC}). The powder data measured after 0.397\,T  FC for  \textcolor[HTML]{8B008B}{$\delta=0.00$} and \textcolor[HTML]{FF8C00}{$\delta=0.03$}  is scaled by a factor of 2 for clarity and taken from \cite{Inazumi1981}.  \\ \textit{Inset}: Magnetization of a \textit{non-stoichiometric} YFe$_2$O$_{4-\delta}$ single crystal, which was grown in a CO$_2$~/~CO-ratio of 2.6$\pm$0.1. (color online)}
\label{fig:mag}
\end{figure}

\begin{figure}[tb]
\setlength\abovecaptionskip{0pt}
\setlength\belowcaptionskip{0pt}
\includegraphics[width=222pt]{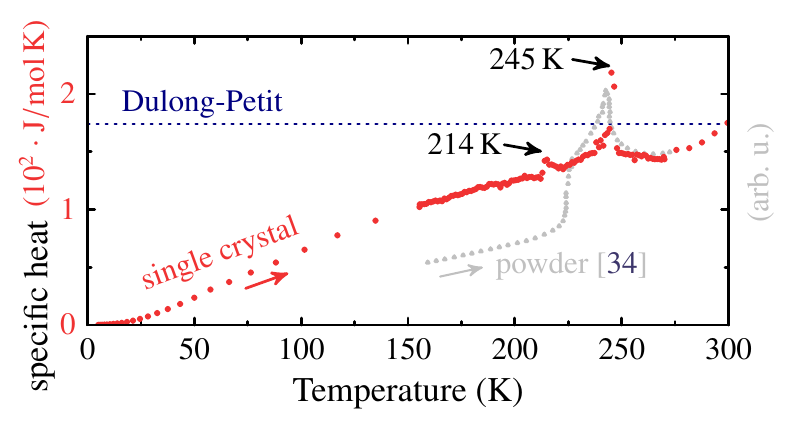}
\caption{Specific heat of a stoichiometric YFe$_2$O$_{4-\delta}$ single crystal, measured during warming. For comparison the powder data from \cite{Tanaka1982} is shown (in arbitrary units). (color online)}
\label{fig:heat_cap}
\end{figure}

The macroscopic magnetization of  YFe$_2$O$_{4-\delta}$ along $\vec{c}_\mathrm{hex}$ is very sensitive to the oxygen off-stoichiometry $\delta$ \cite{Inazumi1981}. Non-stoichiometric samples $(\delta > 0.05)$ show a broad glassy transition around 200\,K (Inset Fig.~\ref{fig:mag}). A previously applied cooling field has a drastic impact on the magnetic moment below this transition. 
This influence is absent in stoichiometric samples $(\delta < 0.03)$ \cite{Inazumi1981}. Stoichiometric YFe$_2$O$_{4-\delta}$ shows two antiferromagnetic transitions with onset at 228.5 and 180\,K on cooling  (Fig.~\ref{fig:mag}). There exists a strong thermal hysteresis on both transitions shifting them 18.5\,K to higher temperatures on warming. The higher temperature transition is also shifted if the sample was only cooled to 200\,K. This implies that both transitions are of first order, which is also confirmed by the presence of latent heat in specific heat measurements (Fig.~\ref{fig:heat_cap}) and is consistent with M\"ossbauer spectroscopy \cite{Tanaka1979}.
In the comparison of the specific heat (Fig.~\ref{fig:heat_cap}) between the single crystal and the polycrystalline sample from \cite{Tanaka1982}, the transitions seem to be more separated for the single crystal. There exists a strong difference at the lower temperature transition with a small peak in the single crystal data and a large continuous increase in the polycrystalline sample.

Away from the transitions the magnetization is essentially linear with applied magnetic field consistent with \cite{Inazumi1981}. It shows neither metamagnetic transitions nor saturation up to 9\,T, in contrast to LuFe$_2$O$_{4-\delta}$~\cite{joost_afm}.

Our single crystals show a magnetization (Fig.~\ref{fig:mag}) comparable to highly stoichiometric powder samples~\cite{Inazumi1981}. The oxygen deficit can be estimated from comparison with the magnetization data on polycrystalline samples from Inazumi \textit{et\,al.}\ \cite{Inazumi1981}, who have determined the deficit by thermogravimetric analysis. Their magnetization data is reproduced in Fig.~\ref{fig:mag} (thin lines).
The magnetization of our single crystals is similar to those of the polycrystalline samples with $\delta = 0.00$ \cite{Inazumi1981}. It shows significantly less ferrimagnetic contamination than those with $\delta = 0.03$, the second most stoichiometric composition in \cite{Inazumi1981}. The magnetization of our stoichiometric single crystals is strongly anisotropic in macroscopic measurements, with $\vec{c}_\mathrm{hex}$ as the Ising-easy-axis. This is also the case for non-stoichiometric YFe$_2$O$_{4-\delta}$ \cite{Sugihara1978} and is also observed in the other RFe$_2$O$_4$ systems \cite{IKEDA1995,IIDA1987}.

\section{Single crystal x-ray diffraction}
\begin{figure}[tb]
\setlength\abovecaptionskip{0pt}
\setlength\belowcaptionskip{0pt}
 {\phantomsubcaption\label{fig:cells}}
 {\phantomsubcaption\label{fig:laue}}
 {\phantomsubcaption\label{fig:recipscan}}
\includegraphics[width=215px]{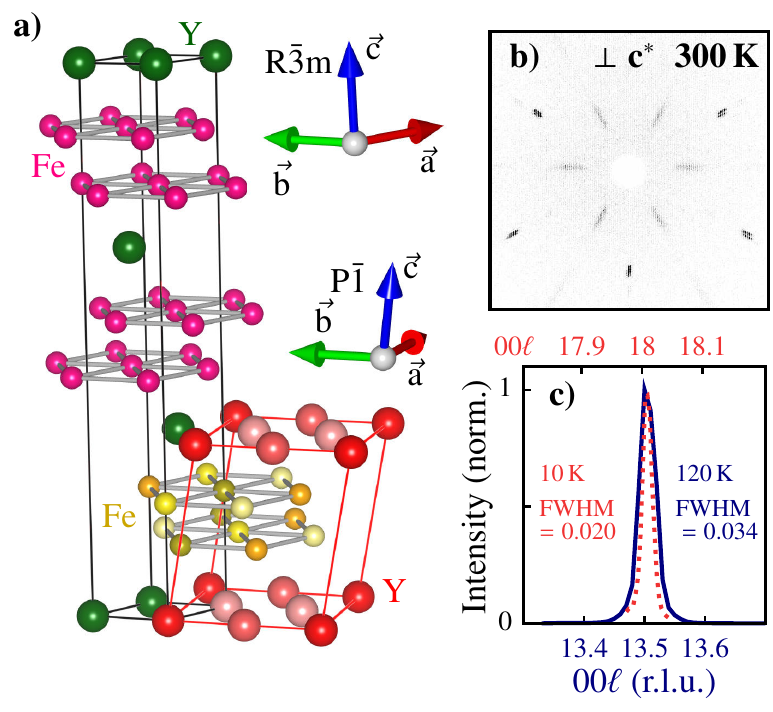}
\caption{a) Crystal structures (oxygen ions are omitted) and unit cells for the R$\bar{3}$m room temperature structure and the distorted P$\bar{1}$ for ($\frac{1}{4}  \frac{1}{4} \frac{3}{4}$) as propagation vector~\cite{isodistort2}. (created with \cite{Momma:db5098});
 b) Laue-diffraction image of YFe$_2$O$_{4-\delta}$ single crystal at 300\,K.
 c) Scan through $(\frac{1}{2}  \frac{1}{2} \textrm{\footnotesize{13.5}})$ along $[00\ell]$ at 120\,K and  $(\textrm{\footnotesize {0}}\,\textrm{\footnotesize {0}}\,\textrm{\footnotesize {18}})$ at 10\,K. The $\ell$-position of the 120\,K peak was corrected, based on the 10\,K UB-matrix. 
  (color online)}
\end{figure}

Like the magnetization, the charge ordering in YFe$_2$O$_{4-\delta}$ was also found to be sensitive to oxygen off-stoichiometry \cite{Mori2008,Horibe2009}. Our in-house x-ray diffraction experiments were performed on a Rigaku SuperNova diffractometer using Cu-$K_\alpha$ radiation. The room temperature reflections (Fig.~\ref{fig:laue}) can be well described with the R$\overline{3}$m structure (Fig.~\ref{fig:cells}), which we will use below to index superstructure reflections. In addition to these reflections we observed diffuse scattering along $(\frac{1}{3} \frac{1}{3}  \ell) $ similar to LuFe$_2$O$_{4-\delta}$ above the charge ordering temperature \cite{Angst2008,joost_co}. The diffuse streaks appear because the charge ordering  occurs first in-plane due to the stronger interactions, which leads to a random stacking of at least medium-range ordered layers along $\vec{c}_\mathrm{hex}$ \cite{joost_co}. Diffuse streaks at room temperature were observed in both stoichiometric and non-stoichiometric YFe$_2$O$_{4-\delta}$ samples. In non-stoichiometric samples, the diffuse streaks remain so upon cooling, in agreement with electron diffraction observations \cite{Horibe2009}. However, in stoichiometric single crystals, the diffuse scattering is replaced by sharp superstructure reflections, upon cooling below the magnetic transition at 228.5\,K. 
Figure~\ref{fig:recipscan} shows reciprocal space scans along $[00\ell]$ through superstructure reflection $(\frac{1}{2} \frac{1}{2} \textrm{\small{13.5}})$ at 120\,K, and through  $(\textrm{\small{0}}\,\textrm{\small{0}}\,\textrm{\small{18}})$ at 10\,K measured at the P09 beamline at PETRA III.
The peaks at 120\,K can be indexed using a single propagation vector  ($\frac{1}{4}  \frac{1}{4} \frac{3}{4}$), if second harmonics and symmetry equivalent vectors are considered. Starting from a distortion of the room temperature structure, assuming a single active $k$-vector and a single order parameter, group theory limits the symmetry to spacegroup P$\overline{1}$ \cite{isodistort2}. Figure~\ref{fig:cells} shows one of the two possible P$\overline{1}$ unit cells in comparison to the R$\overline{3}$m cell. The refinement is complicated, due to six possible twin components together with the low symmetry. To distinguish the Fe$^{2+}$ and Fe$^{3+}$ sites, bond valence sum analysis is required, such as was performed in \cite{joost_co}. This is ongoing work, but beyond the scope of this article and will be presented elsewhere. 

 The peak width of the $(\frac{1}{2} \frac{1}{2} \textrm{\small{13.5}})$ superstructure reflection along (00$\ell$) in Figure~\ref*{fig:recipscan}, corrected for the width of the structural reflection, to take into account instrumental resolution and mosaicity, corresponds to a correlation length of 22 unit cells along $\vec{c}$. This corresponds to a correlation over 66 Fe bilayers, which is larger than typically observed in LuFe$_2$O$_{4-\delta}$ 7~\cite{Park2009} or 9~\cite{Wen2009}. 

Horibe \textit{et\,al.}\ \cite{Horibe2005} found in polycrystalline YFe$_2$O$_{4-\delta}$ with $\delta \leq 0.005$ a coexistence on a mesoscopic scale of two distinct phases at room temperature. One shows diffuse streaks along  $(\frac{1}{3} \frac{1}{3}  \ell) $ attributed to charge ordering ~\cite{Horibe2005}. Matsui \textit{et\,al.}\ \cite{MATSUI1980} and Horibe \textit{et\,al.}\ \cite{Horibe2009} reported $(\frac{1}{3}  \frac{1}{3}  \frac{3}{2})$-type superstructure reflections observed by electron diffraction at room temperature, which according to \cite{Mori2008} disappear upon heating above 368\,K  and reappear after cooling back to room temperature.  Horibe \textit{et\,al.}\ \cite{Horibe2005} reported that these reflections are present for samples with magnetization curves matching those for $\delta=0.00$ of Ref. \cite{Inazumi1981} (see Fig.~\ref{fig:mag}) and with more complex superstructure reflection patterns at lower temperatures. Both magnetic behavior and low-temperature superstructures suggest that these powder samples have an oxygen stoichiometry very similar to the one of our single crystal. The different behavior at room temperature is therefore surprising and suggests that the $(\frac{1}{3}  \frac{1}{3}  \frac{3}{2})$-type modulation reported in \cite{MATSUI1980,Horibe2005} is even more fragile than the low-temperature behavior. It is unclear whether it is a charge order modulation --- the fact that M\"ossbauer spectroscopy at room temperature is virtually the same for stoichiometric and non-stoichiometric YFe$_2$O$_{4-\delta}$ \cite{Tanaka1979,Sugihara1976} hints at a different origin. One possibility is that the formation of YFe$_2$O$_{4-\delta}$ with oxygen excess ($\delta$ $<$ 0) ordering leads to an additional super-structure, as suggested for LuFe$_2$O$_{4-\delta}$ \cite{Bourgeois2012,Hervieu2013}.

Post-annealing of YFe$_2$O$_{4-\delta}$ to tune the oxygen stoichiometry should be possible, as was done for  LuFe$_2$O$_{4-\delta}$ \cite{Hervieu2014}, although it has to be performed at low temperatures to avoid the decomposition of the metastable state. It was tried in the past \cite{Tanaka1982}, but no stoichiometric crystals were obtained. Our attempts always broke the crystals. The magnetization of the resulting parts clearly indicates non-stoichiometric YFe$_2$O$_{4-\delta}$; in comparison with the unannealed crystal, it is improved in regard of stoichiometry. Also, it is difficult to establish a homogeneous oxygen distribution over an annealed macroscopic crystal. After annealing for 40\,h at 400\,$^\circ$C in air, a weak $(\frac{1}{3}\frac{1}{3} 0)$-superstructure, not present before, appears at room temperature in the non-stoichiometric crystal. The magnetization of this sample after annealing does not correspond to stoichiometric YFe$_2$O$_{4-\delta}$. Therefore this superstructure is likely related to ordering of oxygen excess or oxygen defects, and not an intrinsic feature of stoichiometric YFe$_2$O$_{4-\delta}$.

\section{Resonant x-ray diffraction at the Fe K-edge}
\begin{figure}[tb]
\setlength\abovecaptionskip{3pt}
\setlength\belowcaptionskip{0pt}
\includegraphics[width=220px]{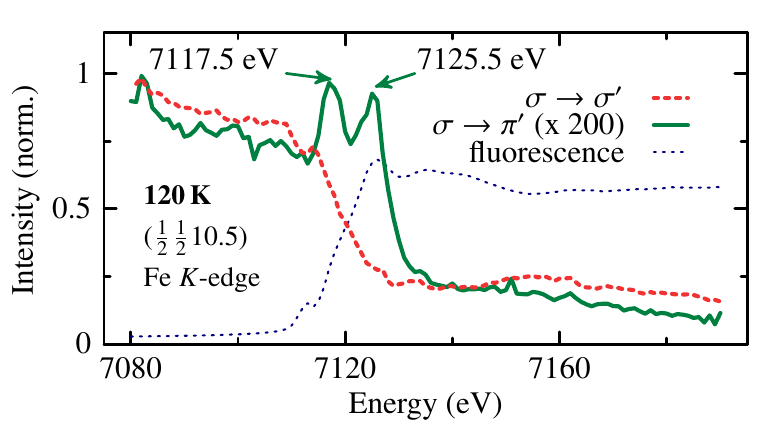}
\caption{Energy dependence of the $(\frac{1}{2}  \frac{1}{2} \textrm{\footnotesize{10.5}})$ superstructure reflection at 120\,K over the Fe $K$-edge, with two resonant features visible in the $\sigma \rightarrow \pi'$ channel at 7115.5\,eV and 7125.5\,eV. (color online)}
\label{fig:energy_120K}
\end{figure}

To examine the origin of the observed low-temperature superstructure reflections we performed a resonant x-ray diffraction experiment at the Fe $K$-edge. The experiment was performed at the P09 beamline \cite{Strempfer:ie5091,Francoual2013} at PETRA III. The resonant features in the spectrum in Figure~\ref{fig:energy_120K} suggest some involvement of Fe-ions in the ordering, which would be consistent with charge ordering of Fe$^{2.5 \pm x}$, where the amount of charge separation is still in debate  in the related LuFe$_2$O$_{4-\delta}$ \cite{lafuerza2014,lafuerza2014_determ,Mulders2009} and other ferrites \cite{Angst2007}. 
In the trigonal bipyramidal crystal field the Fe 3d orbitals are split to (d$_{xy}$,d$_{x^2-y^2}$) and (d$_{yz}$,d$_{xz}$) doublets and a (d$_{z^2}$) singlet \cite{Nagano2007_jpcm}. In Fe$^{3+}$ all five 3d orbitals are  half-filled. In Fe$^{2+}$ the additional electron pairs with one electron  either in the d$_{xy}$ or the d$_{x^2-y^2}$ orbital, leading to a degenerate orbital degree of freedom \cite{Nagano2007_jpcm}.
This gives the possibility of orbital ordering, which is theoretically expected in the RFe$_2$O$_4$ system (R = rare earth) \cite{Nagano2007_jpcm,Nagano2007_prl}.  For LuFe$_2$O$_{4-\delta}$ orbital order was excluded for the ferrimagnetic phase, since there exists an orbital magnetic moment~\cite{joost_co}. For the antiferromagnetic and paramagnetic phases, orbital order was also excluded, because the transitions between them and the ferrimagnetic phase leaves the crystal structure unaffected~\cite{joost_co}. In agreement with this no anisotropies were observed in resonant x-ray diffraction \cite{Mulders2009}.
Our YFe$_2$O$_{4-\delta}$ crystals show a strong Ising-anisotropy in macroscopic magnetization measurements, similar to LuFe$_2$O$_{4-\delta}$~\cite{joost_afm}, which can be explained by an out-of-plane orbital magnetic moment and spin orbit coupling~\cite{Angst2013}.
We probed the possible existence of orbital order in YFe$_2$O$_{4-\delta}$ by polarization analysis on resonant features at the Fe $K$-edge. The polarization analysis follows the procedure described by Mazzoli \textit{et\,al.}\ \cite{Mazzoli2007}. The polarization state of an x-ray beam can be described by Poincar\'{e} Stokes parameters $P_{1} = \frac{|\vec{\epsilon}_{\sigma}|^2-|\vec{\epsilon}_{\pi}|^2}{|\vec{\epsilon}_{\sigma}|^2+|\vec{\epsilon}_{\pi}|^2} $ and $P_{2} = 2\operatorname{Re}\frac{\vec{\epsilon}_{\sigma}^{*}\vec{\epsilon}_{\pi}^{*}}{|\vec{\epsilon}_{\sigma}|^2+|\vec{\epsilon}_{\pi}|^2} $ with the components of the polarization vector $\vec{\epsilon}$ perpendicular $(\sigma)$ or parallel  $(\pi)$ to the scattering plane \cite{Mazzoli2007}. 

\begin{figure}[tb]
\setlength\abovecaptionskip{5pt}
\setlength\belowcaptionskip{0pt}
\includegraphics[width=215px]{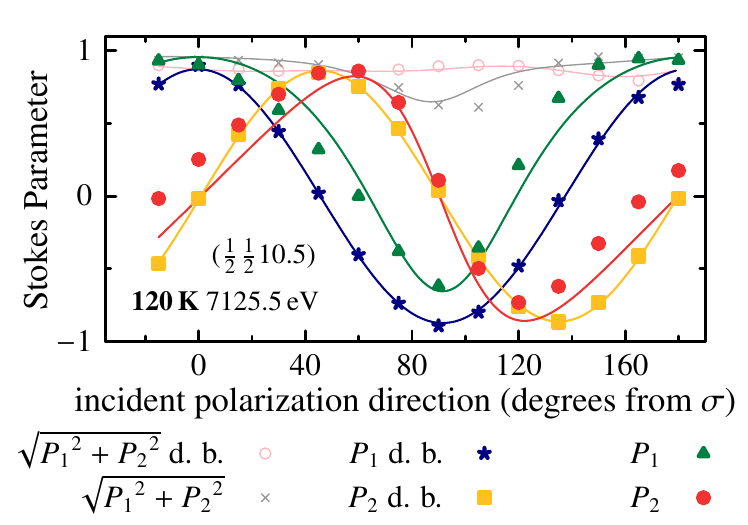}
\caption{Polarization analysis at 7125.5\,eV on $(\frac{1}{2}  \frac{1}{2} \textrm{\footnotesize{10.5}})$\ at 120\,K, and of the direct beam. Solid lines are for the direct beam sinusoidal fits. The solid red and green lines are the expected behavior for an ideal Thomson scatterer. (color online)}
\label{fig:p09_pol_7125}
\end{figure}

Figure~\ref{fig:energy_120K} shows a spectrum over the Fe $K$-edge of the $(\frac{1}{2}  \frac{1}{2} \textrm{\small{10.5}})$-reflection at 120\,K, in $\sigma \rightarrow \sigma'$ and $\sigma \rightarrow \pi'$ polarization channels. In the  $\sigma \rightarrow \pi'$-channel two resonant features are observed at 7117.5 and 7125.5\,eV. Figure~\ref{fig:p09_pol_7125} shows the Stokes parameters of the direct beam and those of the resonant feature at 7125.5\,eV. The solid lines are sinusoidal fits for the direct beam and the calculated ones for an ideal Thomson scatterer in respect to the fits of the  direct beam following \cite{Detlefs2012}. The deviation of $\sqrt{{P_1}^2+{P_2}^2}$ from the ideal amplitude of one arises from the imperfect polarization of the incident beam.

The polarization analysis at 7125.5~eV (Fig.~\ref{fig:p09_pol_7125}) shows only small deviations from the ideal Thomson behavior, indicating a small anisotropic contribution to the resonant diffraction signal, e.g. due to orbital ordering. For the 7117.5\,eV feature the deviations are even smaller.
Both resonant features in $\sigma \rightarrow \pi'$  are weak compared to the intensity in  $\sigma \rightarrow \sigma'$ in Figure~\ref{fig:energy_120K}. Through this the polarization analysis mainly probes the structure and therefore structural anisotropies. In contrast to LuFe$_2$O$_{4-\delta}$~\cite{joost_co,Mulders2009}, there are small anisotropic contributions to the resonant x-ray scattering in YFe$_2$O$_{4-\delta}$, but they remain weak. 

\section{Conclusion and Outlook} 
\label{sec:conclusion}
In summary, we have grown stoichiometric single crystals of YFe$_2$O$_{4-\delta}$ by the optical floating zone method, showing for the first time the two step antiferromagnetic transition observed in highly stoichiometric powder samples $(\delta=0.00)$. The charge ordering observed by x-ray diffraction differs strongly from LuFe$_2$O$_{4-\delta}$, which is reasonable considering the much larger ionic radius of the Y ion. 
It also differs from some of the observations in electron diffraction \cite{Horibe2009} on YFe$_2$O$_{4-\delta}$, suggesting, that the differentiation in 3D and 2D charge ordering at room temperature, belonging to stoichiometric and non-stoichiometric samples respectively, may not be enough to describe the variety of charge order states in YFe$_2$O$_{4-\delta}$. Samples that are considered stoichiometric in regard to the magnetic behavior can obey different charge ordered structures. The magnetization reacts on changes of the oxygen stoichiometry of the order of 0.1\% \cite{Inazumi1981}, the charge ordering seems to be even more sensitive, a direct determination of the absolute oxygen stoichiometry with this precision is difficult. Given the strong competition between different charge order instabilities \cite{Ikeda2002,Ikeda2003} such a strong response to small perturbations may not be too surprising.
The small resonance at the Fe K-edge is a hint for the participation of the Fe ions in the ordering process. Since the resonances are weak and show a dominantly isotropic behavior, any orbital contribution would be small.

Considering what was done on LuFe$_2$O$_{4-\delta}$~\cite{Angst2013}, the availability of stoichiometric single crystals of YFe$_2$O$_{4-\delta}$ offers many new possibilities on the route to understand the RFe$_2$O$_4$ system, such as the refinement of charge order and magnetic structures. Corresponding studies are in progress.

We gratefully acknowledge the support from the initiative and networking fund of the Helmholtz Association of German Research Centers by funding the Helmholtz-University Young Investigator Group ``Complex Ordering Phenomena in Multifunctional Oxides''.
Parts of this research were carried out at the light source PETRA III at DESY, a member of the Helmholtz Association (HGF). 
\bibliographystyle{apsrev4-1}
\bibliography{YFe2O4_bibliography}

\end{document}